\documentclass[]{jfm}
\usepackage{graphicx}
\usepackage{newtxtext}
\usepackage{newtxmath}
\usepackage{natbib}
\usepackage{hyperref}
\usepackage{ulem}
\usepackage{amsmath}
\usepackage[dvipsnames]{xcolor}
\hypersetup{
    colorlinks = true,
    urlcolor   = blue,
    citecolor  = black,
}

\newcommand{\RomanNumeralCaps}[1]
\linenumbers

\title[Impulsive impact of a twin hull]
      {Impulsive impact of a twin hull}

\author{B.-Y. Ni\aff{1},
 \and Y.A. Semenov\aff{1}
 \corresp{\email{yuriy.a.semenov@gmail.com}}}
\affiliation{\aff{1}College of Shipbuilding Engineering, Harbin Engineering University, Harbin, 150001, China}

\begin{document}
\maketitle

\begin{abstract}
An impulsively starting motion of two cylindrical bodies floating on a free liquid surface is considered. The shape of the cross-section of each body and the distance between them is arbitrary. The integral hodograph method is advanced to derive the complex velocity potential defined in a rectangle parameter region in terms of the elliptic quasi-doubly periodic Jacobi theta functions. A system of singular integral equations in the velocity magnitude on the free surface and in the slope of the wetted part of each body is derived using the kinematic boundary condition, which is then solved numerically. The velocity field, the pressure impulse on the bodies, and the added mass coefficients of each body immediately after the impact are determined in a wide range of distances between the bodies and for the cross-sectional shapes such as flat plate and half-circle.
\end{abstract}

\section{Introduction}
A twin hull is the most common design of high-speed semi-displaced vessels, such as catamarans, air cushion vessels, and skeg-type hovercrafts, in which skegs can also serve as demihulls, \cite{Falt05}. A significant interest in the development of these types of vessels has arisen in the last decades due to the need for fast sea transportation of passengers and goods and for the military applications. The advantages of twin hull vessels are mainly due to a large deck area, favorable stability characteristics, seakeeping, and relative fuel saving. At the same time, twin hull vessels may experience slamming loads larger than conventional monohulls due to their smaller draft and the slamming of the bridge, or the deck referred in the literature as "wetdeck slamming".

A comprehensive review on ship slamming and impact load evaluation was recently provided by \cite{Dias and Ghidaglia} . Following \cite{Lafeber_2012}, they explained the water impact of a steep wave as a combination of elementary loading processes consisting of the direct impact of the wave crest, the main liquid body impact with capturing a gas bubble, and the compression/expansion stage due to the bubble oscillations. Based on experimental observations, they concluded that the largest impact loading is associated with a rapid change of the total momentum of the liquid, which can be accounted for using the pressure impulse concept (\cite{Cooker_1995}).

	The pressure impulse concept has a long history, which goes back to the work of \cite{Lagrange1783}, \cite{Joukovskii_1884}, and \cite{HAV_1927}. A remarkable advancement in the impulse impact theory was made by \cite{Karman} with application to seaplane landing. For ship slamming applications, \cite{Wagner} in his foundational paper accounted for the rise of the free surface using the von Karman solution. This approach received further development for solving modern water impact problems (\cite{HowOckOliver04}). There is also a large body of research dealing with the impulse impact concept, which includes the impulsive motion of a body initially floating on a flat free surface (\cite{Iafr_Kor2005}) and dam-break flows (\cite{Korobkin_Yilmaz}).

According to this concept, the pressure at the time of impact tends to infinity, but the impact lasts an infinitesimal
period of time. Recently, Speirs et al. (2021) investigated the impulsive impact of a flat bottom cylinder experimentally accounting for the small compressibility of water. Not only did they found high pressures similar to previous studies, but they also revealed a decrease in the local pressure sufficient to cavitate the liquid. This occurs due to the pressure wave reflecting from the free surface and forming a negative pressure region. Thus, the pressure impulse concept is applicable to the determination of pressure loads for an impact duration small enough to neglect the motion of free surfaces, but large enough to neglect the water compressibility.

All the studies mentioned above consider a single body floating on a free surface. By contrast, our research examines two bodies with a gap between them, which are synchronously set in motion. This case directly corresponds to the slamming of various vessels with twin hulls, such as catamarans.

Our solution method is based on a further extension of the integral hodograph method for solving two-dimensional boundary-value problems with mixed boundary conditions: there are two parts of the flow boundary with free surfaces and two parts corresponding to the two solid bodies where the impermeability condition is implied. The key step of the method is finding the two governing functions: the complex velocity and the derivative of the complex potential, both defined in an auxiliary parameter region. For the determination of the complex velocity, we derived an integral formula for solving a mixed boundary-value problem for an analytical function defined in a rectangular auxiliary parameter region. This formula makes it easy to determine an analytical function from the values of its argument given on the two horizontal sides of the rectangle corresponding to the solid bodies and its modulus given on the two vertical sides corresponding to the two parts of the free surface.

The system of integral equations in the velocity angle along the body and the velocity magnitude along the free surface was derived by employing the kinematic boundary conditions on the body and the free surface. These integral equations were solved numerically to complete the solution.	

The coefficients of the added masses, the pressure impulse acting on the bodies, and the velocity on the free surface including the gap between the bodies were determined for various widths of the gap and for various cross-sectional shapes of the impacting bodies, such as flat plates and half-circles.

\section{Boundary-value problem} \label{sec:2}
\begin{figure}
\centering
\includegraphics[scale=0.65]{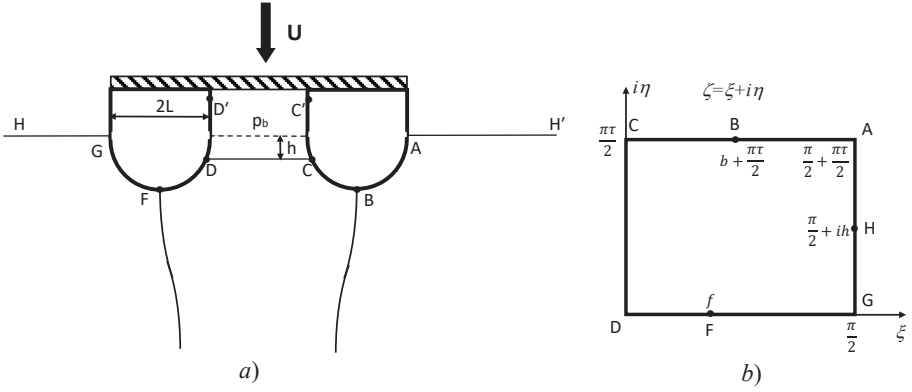}\\
\caption{($a$) definition sketch of the physical domain and ($b$) the parameter, or $\zeta-$plane.}
\label{figure1}
\end{figure}

A sketch of the physical domain is shown in figure \ref{figure1}($a$). Two bodies partially submerged in a liquid and connected by a deck float on the free surface.  In general, the level of the liquid in the gap between the bodies may be different from the calm water level at infinity if the pressure in the gap is different from the ambient pressure. The shapes of the bodies can be different; therefore, the chosen characteristic length $L$ corresponds to the size of one of them. Before the time of impact, $t = 0$, the body and the liquid are at rest. At time $t = 0+$ the body is suddenly set in motion with acceleration $a$ directed downwards so that during an infinitesimal time interval $\Delta t>0$, the velocity of the body reaches the value $U=a\Delta t$.

The problem of a rigid body moving in a fluid body is kinematically equivalent to the problem of a fluid body moving around a fixed rigid body with acceleration $a$ at infinity. We define a Cartesian coordinate system $XY$ attached to the twin body and a coordinate system $X^\prime Y^\prime$  attached to the calm free surface. Each body is assumed to have an arbitrary shape, which is defined by the slope of the body boundary, $\delta_i=\delta_i(S_{bi})$, as a function of the arc length coordinate $S_{bi}$ along the body $i$, $i=1,2$. The liquid is assumed to be ideal and incompressible, and the flow is irrotational. Gravity and surface tension effects during the impact are ignored.

We introduce complex potentials $W(Z)=\Phi(X,Y)+i\Psi(X,Y)$ and $W^\prime(Z)=\Phi^\prime(X,Y)+i\Psi^\prime(X,Y)$ with $Z=X+iY$ in the coordinate systems $XY$ and $X^\prime Y^\prime$, respectively. By integrating Bernoulli's equation over an infinitesimal time interval $\Delta t \rightarrow 0$
in each system of coordinates, we can obtain the relation between the pressure impulses in these systems of coordinates (\cite{semenov_jfm2021})
\begin{equation}
\label{eqv1}
\Pi^\prime=\int_0^{\Delta t}p^\prime dt=-\rho\Phi^\prime=\Pi + \rho UY=-\rho\Phi + \rho UY.
\end{equation}
Here, $\Pi$ and $\Pi^\prime$ are the pressure impulses in the systems $XY$ and $X^\prime Y^\prime$, respectively; $p^\prime$ is the hydrodynamic pressure and $\rho$ is the density of the liquid.

The added mass coefficients, $\lambda_{22}^{\prime}$ and $\lambda_{21}^{\prime}$ , for the body $A^1$ (contour $DFG$) and body $A^2$ (contour $ABC$) of the bodies can be expressed as follows (\cite{Korotkin2009})
\begin{equation}
\label{eqv3}
\lambda_{21}^{\prime1,2} = -\int_{s_D, s_A}^{s_G, s_C}\phi^\prime(s) cos(n,x) ds,  \qquad   \lambda_{22}^{\prime1,2} = -\int_{s_D, s_A}^{s_G, s_C}\phi^\prime cos(n,y) ds,
\end{equation}
Here, $\phi(s) = \Phi(S)/(LU)$ is the dimensionless potential normalized to $U$ and $L$; $s_A$, $s_C$, $s_D$, $s_G$ are the arc length coordinates of points $A$, $C$, $D$, $G$; $x = X/L$, $y = Y/L$, $s = S/L$; the dimensionless velocity on the free surface is $v = |V|/U$.

By substituting (\ref{eqv1}) in the dimensionless form into (\ref{eqv3}) we obtain the relation between the added mass coefficients in the systems of coordinates $X^\prime Y^\prime$ and $XY$:
\begin{equation}
\label{eqv4}
\lambda_{21}^{\prime1,2} = \lambda_{21}^{1,2}, \qquad  \lambda_{22}^{\prime1,2} = \lambda_{22}^{1,2} - a^{\ast 1,2},
\end{equation}
where
\[
a^{\ast 1,2} = \int_{s_D, s_A}^{s_G, s_C}y(s)cos(n,y)ds.
\]

The coefficient $\lambda_{22}^{1,2}$ accounts for the acceleration of the liquid in the $y$ direction during the impact, which is similar to gravity and causes the buoyancy force.

In the following, the objective is to determine the velocity potential of the flow $\phi(s)$ in the system of coordinates $XY$, in which the liquid suddenly starts to move upward with velocity $U$.

\section{Conformal mapping} \label{sec:3}
We choose the rectangle $DGAC$ in the $\zeta-$plane with the vertexes $(0,0)$, $(\pi/2,0)$, $(\pi/2,\pi\tau/2)$ and $(0,\pi\tau/2)$, respectively (figure \ref{figure1}$b$), as an auxiliary parameter region. Here, $\tau$ is an imaginary number. The horizontal length of the rectangle is equal to $\pi/2$, and its vertical length is equal to $\pi|\tau|/2$. The corresponding points of the rectangle and the flow region are denoted by the same letters. The horizontal sides of the rectangle $DG$ and $AC$ correspond to the surface of body $A^1$ and $A^2$, respectively. The vertical side $GA$ of the rectangle ($\xi =\pi/2, 0<\eta <\pi/2)$ corresponds to the free surface outside the twin body, i.e., $HG$ $(0\le\eta<ih)$ and $AH^\prime$ $(h<\eta\le\pi/2)$. The vertical side $DC$ ($\xi =0$, $0\le \eta \le \pi/2$) corresponds to the free surface between the bodies. The positions of the stagnation points $F$ ($\zeta=f$) and $B$ ($\zeta=b+\pi\tau/2$), and point $H$ ($\zeta=\pi/2+ih$) corresponding to infinity are to be determined from the solution of the problem and physical considerations.

We formulate boundary-value problems for the complex velocity function, $dw/dz$, and for the derivative of the complex potential, $dw/d\zeta$, both defined in the  $\zeta-$plane. If these functions are known, then the derivative of the mapping function can be obtained (\cite{Michell_1890, Joukovskii_1890})
\begin{equation}
\label{eq_dz}
\frac{dz_m}{d\zeta}  = \frac{dw}{d\zeta}/ \frac{dw}{dz},
\end{equation}
and its integration in the $\zeta-$plane gives the mapping function $z = z_m(\zeta)$ relating the coordinates in the parameter and the physical planes.

If a complex function is defined in a rectangle, it can be extended periodically onto the whole complex $\zeta-$ plane using the symmetry principle \cite{Gurevich65}. This periodic domain is consistent with a definition domain of doubly-periodic elliptic functions with periods $\pi$ and $\pi\tau$ along the $\xi$ and the $\eta$ axes, respectively. For example, we can derive expressions for the complex velocity and for the derivative of the complex potential by using Jacobi's theta functions.

\subsection{Expressions for the complex velocity and the derivative of the complex potential} \label{subsec:31}
The body is considered to be fixed; therefore, the velocity direction on each body is determined by the slope of the body, $\delta_i=\delta_i(s_{bi})$, $i=1,2$. Besides, at this stage, we assume that the velocity direction $\chi_i(\xi)=-d \delta_i/ds [s_{bi}(\xi)] $ is a known function of the parameter variable $\xi$ and the velocity magnitude on each part of the free surface is a known function of the parameter variable $\eta$, or $v_i=v_i(\eta)$, $i=1,2$. To solve this boundary-value problem, we have to derive an integral formula that determines a complex function defined in a rectangular domain from its values on the horizontal and vertical sides of the rectangle. By applying the special point method (\cite{Gurevich65}) and proceeding similarly to \cite{semenov_Wu2020}, the following integral formula can be obtained
\begin{eqnarray}
\label{dwdz}
&&\frac{dw}{dz}=\exp\left[-\frac{1}{\pi}\int_0^{\pi/2}\frac{d\chi_1}{d\xi}\ln\frac{\vartheta_1(\zeta-\xi)}{\vartheta_1(\zeta+\xi)} d\xi -\frac{1}{\pi}\int_{\pi/2}^0\frac{d\chi_2}{d\xi}\ln\frac{\vartheta_4(\zeta-\xi)}{\vartheta_4(\zeta+\xi)} d\xi\right.\\ \nonumber
&+&\frac{i}{\pi}\left.\int_0^{\frac{\pi|\tau|}{2}}\frac{d\ln v_1}{d\eta}\ln\frac{\vartheta_2(\zeta-i\eta)}{\vartheta_2(\zeta+i\eta)} d\xi + \frac{i}{\pi}\int_{\frac{\pi|\tau|}{2}}^0 \frac{d\ln v_2}{d\eta}\ln\frac{\vartheta_1(\zeta-i\eta)}{\vartheta_1(\zeta+i\eta)} d\xi + c_1\zeta + c_2 + ic_3\right],
\end{eqnarray}
where $\vartheta_i(\zeta)$, i=1,2,4, are Jacobi's doubly-periodic theta functions, and $c_i$, $i=1,3$, are real constants to be determined from the conditions for the magnitude and direction of the velocity at infinity, $\zeta=\zeta_h=\pi/2+ih$, and the angle determining the orientation between the bodies.
By substituting into (\ref{dwdz}) $\zeta$ laying on the boundary of the rectangle, we can verify that the boundary conditions for the function dw/dz are satisfied:
\[
\arg\left(\frac{dw}{dz}\right)_{\zeta=\xi}=\chi_1(\xi), \qquad \arg\left(\frac{dw}{dz}\right)_{\zeta=\xi+\frac{\pi\tau}{2}}=\chi_2(\xi),
\]
\[
\left|\frac{dw}{dz}\right|_{\zeta=\frac{\pi}{2}+i\eta}=\frac{v_1(\eta)}{v_1(\eta)_{\eta=h}}, \qquad
\left|\frac{dw}{dz}\right|_{\zeta=i\eta}=\frac{v_2(\eta)}{v_2(\eta^\ast)},
\]
where $0 < \eta^\ast < \pi/2$ corresponds to some point on the free surface $DC$.

The velocity direction on the body is determined by the slope of the body and the position of the stagnation point: at the stagnation points $F$ and $B$ the angle of the velocity direction changes by $\pi$; thus we can write
\begin{equation}
\label{bc_dwdz}
\chi_i (\xi ) = \arg \left(\frac{dw}{dz} \right) =
\left\{{\begin{array}{l}
- \delta_i(\xi) + \pi,\quad 0 \leq \xi \leq f, \quad \,\, \eta = 0, \quad\quad \,\, i=1, \\
\qquad \qquad \qquad b \leq \xi \leq \pi/2, \,\, \eta = \pi|\tau|/2,  i=2, \\
- \delta_i(\xi) , \quad\quad \,\,\,\, f \leq \xi \leq \pi/2, \,\, \eta = 0, \quad\quad\,\, i=1, \\
\qquad \qquad \qquad 0 \leq \xi \leq b, \,\, \quad \eta = \pi|\tau|/2, \,\, i=2.
\end{array}} \right.
\end{equation}

By substituting (\ref{bc_dwdz}) into (\ref{dwdz}) and evaluating the integrals over the step change of the functions $\chi_1(\xi)$ and $\chi_2(\xi)$ at points $F$ and $B$, we obtain the following expression for the complex velocity
\begin{eqnarray}
\label{dwdz_final}
&&\frac{dw}{dz}= \frac{\vartheta_1(\zeta-f)\vartheta_4(\zeta-b)}{\vartheta_1(\zeta+f)\vartheta_4(\zeta+b)} \exp\left[-\frac{1}{\pi}\int_0^{\frac{\pi}{2}}\frac{d\delta_1}{d\xi}\ln\frac{\vartheta_1(\zeta-\xi)}{\vartheta_1(\zeta+\xi)} d\xi -\frac{1}{\pi}\int_{\frac{\pi}{2}}^0\frac{d\delta_2}{d\xi}\ln\frac{\vartheta_4(\zeta-\xi)}{\vartheta_4(\zeta+\xi)} d\xi\right. \nonumber \\
&+&\frac{i}{\pi}\left.\int_0^{\frac{\pi|\tau|}{2}}\frac{d\ln v_1}{d\eta}\ln\frac{\vartheta_2(\zeta-i\eta)}{\vartheta_2(\zeta+i\eta)} d\xi + \frac{i}{\pi}\int_{\frac{\pi|\tau|}{2}}^0 \frac{d\ln v_2}{d\eta}\ln\frac{\vartheta_1(\zeta-i\eta)}{\vartheta_1(\zeta+i\eta)} d\xi + c_1\zeta + c_2 + ic_3\right].
\end{eqnarray}

We derive the derivative of the complex potential, $dw/d\zeta$, using Chaplygin's singular point method (\cite{Gurevich65}, Chapter 1, $\S5$). The function $dw/d\zeta$ has simple zeros at the points $\zeta=f$  and $\zeta=b+\pi\tau/2$ that correspond to the stagnation points $F$ and $B$ at which the complex potential splits into two branches. At point $H$ ($\zeta_H=\pi/2+ih$) the complex potential has a pole of the first order corresponding to a half-infinite flow domain (\cite{Gurevich65}, Chapter 1, $\S4$); therefore, the derivative of the complex potential at the point $\zeta_H$  has a pole of the second order. By extending the derivative of the complex potential symmetrically with respect to the sides $DG$ and $DC$ with the aim to provide real values of the complex potential on the vertical and horizontal sides of the rectangle, we have to put singularities of the same order in the symmetric points $\zeta=-f$,  $\zeta=-b+\pi\tau/2$ and $\zeta=\pi/2-ih$. Then, using Liouville's theorem, the derivative of the complex potential can be written in the form
\begin{equation}
\label{dwdzeta}
\frac{dw}{d\zeta}= K\frac{\vartheta_1(\zeta-f)\vartheta_1(\zeta+f)\vartheta_4(\zeta-b)\vartheta_4(\zeta+b)}
{\vartheta_2^2(\zeta-ih)\vartheta_2^2(\zeta+ih)}
\end{equation}
Here, we used the relation between the theta functions $\vartheta_i(\zeta)$, $i=1,2,3,4$; $K$ is a real constant.

Equations (\ref{dwdz_final}) and (\ref{dwdzeta}) include the parameters  $b$, $f$, $h$, $K$, $\tau$, $c_1$, $c_2$ and $c_3$ and the functions  $v_i(\eta)$, $\delta_i(\xi)$, $i=1,2$, all to be determined from physical considerations and the kinematic boundary condition on the free surface, the body boundary.

\subsection{System of equations in the unknowns $b$, $f$, $h$, $K$, $\tau$, $c_1$, $c_2$ and $c_3$.} \label{subsec:32}
The constants $c_1$, $c_2$, and $c_3$ are determined from the following conditions: the velocity direction at point $A$ ($\zeta_A=\pi/2+\pi\tau/2$), $\arg(dw/dz)_{(\zeta=\zeta_A)}=-\delta_A$; the velocity magnitude at infinity, point $H$ ($\zeta=\zeta_H$), is assumed to be equal to unity, or  $|dw/dz|_{(\zeta=\zeta_H)}=1$; the velocity direction at point $G$ ($\zeta_G=\pi/2$) the  argument of the complex velocity $\arg(dw/dz)_{(\zeta=\zeta_G)}=-\delta_G$.

For the determination of the unknowns $b$, $f$, $h$, $K$, and $\tau$ we use the following conditions: the pressure impulse must be zero at contact points points $D$ and $G$, or
\begin{equation}
\label{syseq1}
w_G=\int_0^{\pi/2}\left. \frac{dw}{d\zeta}\right|_{\zeta=\xi}d\xi=0;
\end{equation}
the wetted length of the fist (second) body is $S_{w1}$($S_{w2}$)
\begin{equation}
\label{syseq2}
\int_0^{\pi/2}\left|\frac{dz_m}{d\zeta}\right|_{\zeta=\xi({\xi+\frac{\pi\tau}{2}})}d\xi=S_{w1}\left({S_{w2}}\right);
\end{equation}
the level of the free surface at infinity at left and right hand sides is the same. By integrating the derivative of the mapping function (\ref{eq_dz}) along an infinitesimal semi-circle centred at point $\zeta_H=\pi/2+ih$, we have
\begin{equation}
\label{syseq3}
\Im\left(\oint_{\zeta=\zeta_H}\frac{dz_m}{d\zeta}d\zeta \right)=\Im\left(\frac{i\pi}{-iv_\infty} \begin{array}{c}
~~~  \\
\mbox{res}\\
~^{\zeta=\zeta_H}
\end{array}
\frac{dw}{d\zeta} \right)=
\Im\left\{ \frac{\pi}{-v_\infty} \frac{d}{d\zeta }\left[\frac{dw}{d\zeta}(\zeta-\zeta_H)^2\right] \right\}=0;
\end{equation}
the distance between the bodies,  $S_{gap}$, is obtained by integrating the derivative of the mapping function along the vertical side $DC$ of the rectangle
\begin{equation}
\label{syseq4}
\Re\left(\int_0^{\frac{\pi|\tau|}{2}} \left. \frac{dz_m}{d\zeta} \right|_{\zeta=i\eta}\right) d\eta =S_{gap}
\end{equation}

\subsection{Body boundary conditions for the function $\delta_i(\xi)$, $i=1,2$} \label{subsec:33}
By using the given function of the slope of the bodies, $\delta_(s_{bi})i$, $i=1,2$, where $s_{bi}$ is the arc length coordinate along $i$ body, and changing the variables, we obtain the following integro-differential equations in the functions $\delta_i(\xi)$:
\begin{equation}
\label{delta_xi}
\frac{d\delta_i}{d\xi}=\frac{d\delta_i}{ds}\frac{ds_{bi}}{d\xi}=\frac{d\delta_i}{ds}\left|\frac{dz_m}{d\zeta}\right|_{\zeta=\zeta_i}, \qquad 0 \le \xi \le \pi/2, \qquad  i=1,2,
\end{equation}
where $\zeta_1=\xi$ and $\zeta_2=\xi+\pi\tau/2$ and the derivative of the mapping function (\ref{eq_dz}) is used.

\subsection{Free surface boundary conditions  for the function $v(\eta)$, $0<\eta<\infty$} \label{subsec:34}
An impulsive impact is characterized by an infinitesimally small time interval $\Delta t \rightarrow 0$ such that the position of the free surface does not change during the impact. From the Euler equations it follows that the impact-generated velocity is perpendicular to the free surface:
\begin{equation}
\label{delta_xi}
\arg \left( \left.\frac{dw}{dz}\right|_{\zeta=\overline{\zeta_i}} \right) = -\frac{\pi}{2}, \qquad i=1,2,
\end{equation}
where $\overline{\zeta_1}=\pi/2 +i\eta$, $\overline{\zeta_2}=i\eta$, $0 \le \eta \le \pi/2$.

Taking the argument of the complex velocity from (\ref{dwdz_final}), we obtain the following integral equation in the functions $d \ln v_i/d\eta$, $i=1,2$:
\begin{eqnarray}
\label{inteq_vv}
\frac{(-1)^{i+1}}{\pi}\int_0^{\frac{\pi|\tau|}{2}}\frac{d\ln v_i}{d\eta^\prime}\ln\left|\frac{\vartheta_1(\eta-\eta^\prime)}{\vartheta_1(\eta+\eta^\prime)} \right|d\eta^\prime=-\frac{\pi}{2}-I_{bi}(\eta)-I_{vi}(\eta)-c_1\eta - c_3,\\
0 \le \eta \le \pi|\tau|/2, \quad i=1,2, \nonumber
\end{eqnarray}
where
\[
I_{bi}(\eta)=\frac{1}{\pi}\int_0^{\frac{\pi}{2}}\frac{d\delta_1}{d\xi}\Im\left\{\ln \frac{\vartheta_{3-i}(i\eta-\xi)}{\vartheta_{3-i}(i\eta+\xi)} \right\} d\xi + \frac{1}{\pi}\int_{\frac{\pi}{2}}^0\frac{d\delta_2}{d\xi}\Im\left\{\ln \frac{\vartheta_{i+2}(i\eta-\xi)}{\vartheta_{i+2}(i\eta+\xi)} \right\}d\xi,
\]
\[
I_{vi}(\eta)= \Im\left\{\ln\frac{\vartheta_{3-i}(i\eta-f)\vartheta_{i+2}(i\eta-b)}{\vartheta_{3-i}(i\eta+f)\vartheta_{i+2}(i\eta+b)}  \right\}+
\frac{(-1)^i}{\pi}\int_0^{\frac{\pi|\tau|}{2}}\frac{d\ln v_{3-i}}{d\eta^\prime}\ln\left|\frac{\vartheta_2(i\eta^\prime -\eta)}{\vartheta_2(i\eta^\prime +\eta)}\right|d\eta^\prime.
\]
Equation (\ref{inteq_vv}) is a Fredholm integral equation of the first kind with a logarithmic kernel, which is solved numerically.

By using a small-time expansion, the first order approximation of the free surface can be obtained as follows:
\begin{equation}
\label{free_surf}
\bar{\eta}(x,t)=\bar{\eta}(x,0) + \frac{\partial \bar{\eta}}{\partial t}(x,0)t + \ldots.
\end{equation}
Here, the kinematic boundary condition on the free surface, (the velocity is perpendicular to the free surface) is used:
\[\frac{\partial \bar{\eta}}{\partial t}[x(\eta),0]=-\Im\left( \frac{dw}{dz}\right)_{\zeta=i\eta} =v[x(\eta),0].\]

\section{Results and discussion} \label{sec:4}
\begin{figure}
\centering
\includegraphics[scale=0.5]{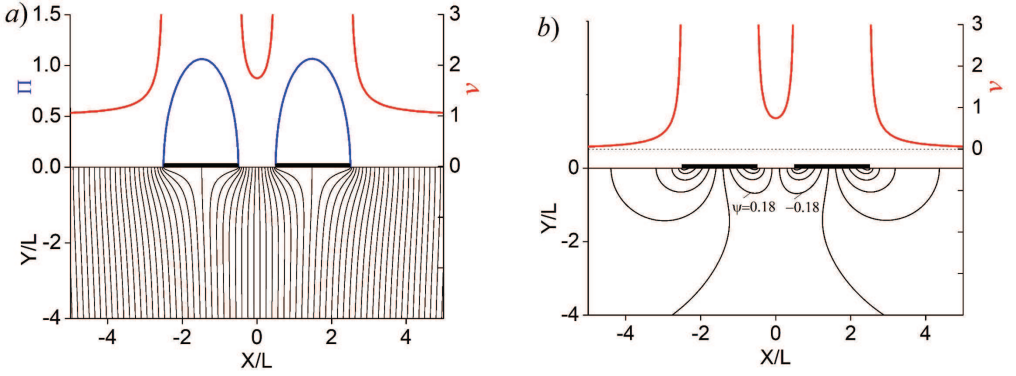}
\caption{The streamlines patterns (left axis), the velocity magnitude on the free surface (right axis, red curve) and the pressure impulse (left axis, blue curve) in the system of coordinates attached to the plate ($a$) and to the liquid at rest ($b$). The length of each plate is $2L$ and the gap $S_{gap}=L$.}
\label{figure:2}
\end{figure}
Figure \ref{figure:2} shows streamlines patterns (left axis) and the velocity magnitude on the free surface right axis) for the impulsive impact of twin plates in the system of coordinates attached to the plate ($a$) and to the liquid at rest ($b$). As expected, the density of the streamlines is higher near the plate edges, which corresponds to a higher velocity. This agrees with the velocity magnitude on the free surface shown in the upper part of the figures. The magnitude of the velocity rapidly increases tending to infinity at the edges. On the free surface between the plates, the velocity reaches its minimum at the middle of the gap (due to the flow symmetry), which is higher than the velocity at infinity since the plates push the liquid into the gap. The pressure impulse on the plate is shown as a blue curve. The pressure impulse is the same in both systems of coordinates (equation \ref{eqv1} ) because the thickness of the plate is zero.
\begin{table}
  \begin{center}
\def~{\hphantom{0}}
  \begin{tabular}{lcccccccc}
$Sgap/L$   &   $\tau$     &  $\lambda_{22}$ & $Q$   & $v_m$   & $v_{av}=Q/S_{gap}$  & $v_{av}/v_m$ & $2b/\pi$ \\[3pt]
$\infty^\ast$    & -      &  $\pi/2$          &   -   & -       & -                  & -            & -      \\[3pt]
   2             & 1.285  & 1.606           & 1.714 & 0.319   & 0.857              & 2.687        & 0.583  \\
   1             & 1.056  & 1.664           & 1.590 & 0.735   & 1.590              & 2.163        & 0.616  \\
   0.5           & 0.880  & 1.746           & 1.455 & 1.540   & 2.910              & 1.890        & 0.649  \\
   0.2           & 0.712  & 1.875           & 1.278 & 3.710   & 6.390              & 1.722        & 0.689  \\
   0.1           & 0.618  & 1.978           & 1.153 & 6.954   & 11.53              & 1.658        & 0.714  \\
   0.015         & 0.450  & 2.228           & 0.879 & 37.53   & 59.37              & 1.582        & 0.767  \\
   0.002         & 0.350  & 2.410           & 0.697 & 220.7   & 348.3              & 1.578        & 0.804  \\
   $2.38\cdot10^{-6}$& 0.2  & 2.697           & 0.395 & 105699. & 165851             & 1.569        & 0.869  \\
   $3.59\cdot10^{-13}$& 0.1 &  2.877          & 0.198 & $3.51\cdot10^{11}$ & $5.51\cdot10^{11}$& 1.569  & 0.923  \\
   $0^\ast$           & -   & $\pi$      & -     &                  &                 &        &        \\
  \end{tabular}
  \caption{Main parameters  of the impulsive impact of the twin plates.}
  \label{tab:1}
  \end{center}
\end{table}

The main flow parameters are shown in Table \ref{tab:1} for various widths of the gap: the added mass coefficient $\lambda_{22}$; the discharge through the gap, $Q$; the minimal velocity in the gap, $v_m$; the average velocity in the gap, $v_{av}$; the ratio $v_{av}/v_m$; the parameter $2b/\pi$. Due to the flow symmetry the parameter $f=b$ and $h=\pi|\tau|/4$.
As the gap tends to infinity, the flow near each plate becomes the same as for an isolated flat plate. As the gap tends to zero, the total length of the plates increases twice, $4L$, and the total added mass increases by a factor of $4$ ($L^2$) and becomes $2\pi$.  It can be seen that the added mass of each plate approaches $\pi$, or the total added mass of the two plates approaches $2\pi$.  As the gap getting smaller, the velocity in the gap increases due to infinite velocity at the edges of the plate. This causes a moderate flow rate through the gap even for extremely  small gap. The ratio of the average velocity to the minimal velocity in the gap tends to the value which is close to $\pi/2$.

\begin{figure}
\centering
\includegraphics[scale=0.5]{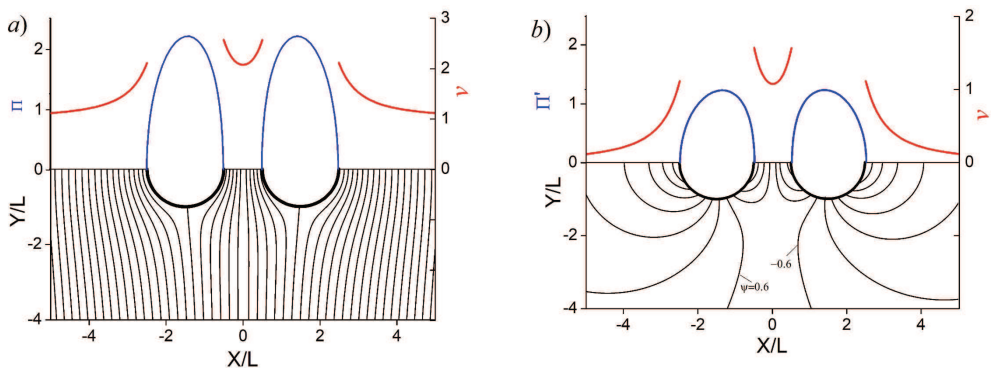}
\caption{Same as in figure 2 but for two half-circles.}
\label{figure:3}
\end{figure}
Figure \ref{figure:3} shows the results for twin half-circles of diameter $2L$ each and gap $S_{gap} = L$. The streamlines corresponding to the stagnation points in figure \ref{figure:3}$a$ are slightly inclined to each other so that the distance between them increases near the bodies. This means that the velocity there is smaller than at infinity, but then it rapidly increases in the gap near the free surface. The velocity on the free surface is finite everywhere including the contact points of the free surface and the circles. This is because the slope of the circles at the contact points equals $\pi/2$, which is the same as the velocity direction on the free surface, and therefore there is no jump in the velocity direction at the contact points, and, correspondingly, there is no singularity  in the function of the complex velocity at these points.

	The streamlines corresponding to the impulsive motion of the twin half-circles are shown in figure \ref{figure:3}$b$ together with the velocity distribution on the free surface in the upper part of the figure. All the streamlines start on the circles, which move down with velocity $U$, and end on the free surface, except for the two streamlines separating those ending on the free surface in the gap between the half-circles and on the rest of the free surface. The half-circles push the liquid into the gap, which results in a much higher velocity on the free surface in the gap than on the rest of the free surface.

\begin{table}
  \begin{center}
\def~{\hphantom{0}}
  \begin{tabular}{lcccccccc}
$Sgap/L$         & $\tau$ & $\lambda_{22}$ & $\lambda_{21}$ & $Q$ & $v_m$ & $v_{av}=Q/S_{gap}$  & $v_{av}/v_m$ \\[3pt]
$\infty^\ast$    & -      &  $\pi/2$       &   -    & -       & -                  & -          & -      \\[3pt]
   5.80          & 1.3    & 1.588          &  0.016 & 1.737   & 0.136              & 0.299      & 2.202   \\
   2.35          & 0.9    & 1.712          &  0.047 & 1.5     & 0.454              & 0.638      & 1.406  \\
   1.33          & 0.7    & 1.850          &  0.098 & 1.327   & 0.827              & 0.998      & 1.206  \\
   0.648         & 0.5    & 2.082          &  0.212 & 1.080   & 1.542              & 1.667      & 1.081  \\
   0.225         & 0.3    & 2.458          &  0.463 & 0.747   & 3.228              & 3.320      & 1.029  \\
   0.099         & 0.1    & 2.715          &  0.653 & 0.535   & 5.335              & 5.386      & 1.010  \\
  \end{tabular}
  \caption{Main parameters  of the impulsive impact of the twin half-circles. }
  \label{tab:2}
  \end{center}
\end{table}

The main flow parameters are shown in Table \ref{tab:2} for various widths of the gap. Due to the identical geometry of the bodies, there is a line of flow symmetry, $x=0$, which can be considered as a rigid wall,  which results in a nonzero force acting on each body in the $x-$direction; therefore, the added mass coefficient $\lambda_{21}\neq 0$. As the gap decreases, the velocity on the free surface increases, but remains finite in contrast to the results corresponding to the flat plate. The ratio of the average velocity to the minimal velocity in the gap approaches unity; this means that the velocity distribution on the free surface in the gap becomes almost uniform.

\section{Conclusions} \label{sec:6}
An impulsively starting flow generated by a pair of bodies floating on a free surface is studied using the integral hodograph method. A rectangle is chosen as the parameter plane, and the solution is obtained in terms of Jacobi's quasi-doubly periodic  theta functions. The boundary-value problem is reduced to a system of integral equations in the functions of the velocity direction on the solid boundary and the velocity magnitude on the free surface, which are solved numerically.

The streamline patterns, the velocity distribution on the free surface, and the pressure impulse along the body are determined for various cross-sectional shapes such as plates and half-circles. Although the results are shown for the identical bodies, the solution allows the shape of the bodies to be different.

The main flow parameters are determined as a function of the width of the gap. It is shown that as the gap tends to infinity, the flow parameters tend to their values corresponding to the impulsive impact of a single body; as the gap tends to zero, the body becomes a single body, but with a larger dimension.

The obtained solution is also applicable to the study of an upward impact since it generates an identical magnitude of the velocity on the free surface. However, the velocity direction is opposite. The obtained solution can be considered as a first-order solution in solving the problem using the method of small time series.

\section*{Acknowledgements} This work is supported by the National Natural Science Foundation of China (Nos.52192693, 52192690 and U20A20327), to which the authors are most grateful.

\section*{Declaration of interests} The authors report no conflict of interests.

\end{document}